 \documentstyle[aps]{revtex}  

\begin{document}
\draft

\title{Evaporation of a packet of quantized vorticity}

\author{Carlo F. Barenghi and David C. Samuels}
\address{Mathematics Department, University of Newcastle,
Newcastle NE1 7RU, UK}
\date{\today}
\maketitle

\begin{abstract}
A recent experiment has confirmed the existence of quantized turbulence
in superfluid $^3$He-B and suggested that turbulence is
inhomogenous and spreads away from the region around the vibrating wire
where it is created. To interpret the experiment
we study numerically the diffusion of a packet of quantized
vortex lines which is initially confined inside a small region of space.
We find that reconnections fragment the packet into a gas of small
vortex loops which fly away. We determine
the time scale of the process and find that it is
in order of magnitude agreement with the experiment.
\end{abstract}


The decay of superfluid turbulence at very low temperatures
raises the fundamental question of the existence near absolute zero
of an energy cascade from large to small scales. In the case of
ordinary turbulence, the nonlinear terms of the Navier - Stokes equation
distribute the energy over various scales  without changing the
total energy. The process leads to Richardson's cascade of bigger eddies
breaking up into smaller eddies, until the wavenumber 
is large enough that kinetic energy is dissipated by viscous forces.
In the case of superfluid turbulence, recent work indicates
that generation of sound plays the role of 'sink' of kinetic 
energy\cite{sound}. Furthermore
it appears that the generation of small scales occurs via creation
of helical waves\cite{Kelvin} of higher and higher
wavenumbers on the quantized vortex filaments (Kelvin waves), and via
creation of small vortex loops\cite{Tsubota}.
The nonlinear mechanism behind both
processes is vortex reconnection, either indirectly (reconnections
create cusps which relax into large amplitude Kelvin waves)
or directly (reconnections create small vortex loops).

The aim of this Letter is to show that the formation of small vortex
loops is particularly important if the turbulence
is not homogeneous, a case which is less
investigated than homogeneous turbulence but is relevant to
experiments performed at the lowest temperatures.
In the case of $^4$He many experiments\cite{He4} have been performed
above $1.3\rm K$ but much less is known about
lower temperatures. In the most relevant study\cite{Davis}
turbulence was produced by oscillating
a grid at temperatures as low as $20 \rm mK$ ($0.01 T_c$ where $T_c$ is
the critical temperature). Although there is no direct evidence, it is
reasonable to expect that the vortex tangle was localized
in the region of the grid. The detection was based on trapping
ions on the quantized vortices, and measurements of the collected charge
indicated that the total amount of turbulence in the cell decayed in time.
In the case of $^3$He-B, direct studies of vortices are typically
done in a rotating cryostat\cite{Krusius}, and an indirect observation of
turbulence\cite{Pickett}\cite{Bradley} has been confirmed only
recently\cite{Fischer}.
In this experiment turbulence was created by vibrating a small wire
shaped like a half-circle at temperatures around $0.11 T_c$. Again, we
expect that turbulence was localized in the region of the vibrating
wire. Additional wires were used to detect the Andreev reflection of
quasi-particles from the turbulent superfluid velocity field.
The experimental set-up could not tell unambiguously whether
the growth time of the screening reflects the intrisic temporal decay
of the vorticity or the spatial evolution of the vortex tangle
away from the region in which it was created. Nevertheless,
the authors argued from the temperature
dependence of the effect that it is more likely that a spatial
evolution of the vortex tangle took place; typically, the spatial
and temporal scales were $1~\rm mm$ and $1~\rm sec$
respectively.

In both experiments, the lack of flow visualization 
makes the interpretation of the data difficult, so 
numerical simulations can give insight into the problem. Thus
motivated, the simple question which we address 
is: what is the fate of a tangle of quantized vortex loops
initially confined in a small region? Can any physical processes cause
the packet of
quantized vorticity to diffuse out and spread in space? 

In a classical Navier - Stokes fluid, diffusion of vorticity is caused
by viscous forces. In the absence of viscosity (a
perfect Euler fluid) vortex reconnections are not possible and the
initial topology is frozen into the fluid; vortex loops distort
but remain linked to each other, so conservation of helicity
prevents the packet from spreading\cite{Moffatt}.
In a superfluid, vortex
reconnections are possible\cite{Koplik}, so the question which we ask
is whether the packet of quantized vorticity remains localized in the
initial region or not, and if not, how fast it diffuses away.
Additional motivation 
arises from the recent result\cite{complexity} that a vortex tangle's
geometrical and topological complexity (the amount of twists and links
of the turbulent  structure) seems to be related to its ability to
diffuse in the time scale under consideration.

We represent a quantized vortex filament as a
space curve ${\bf s}={\bf s}(\xi,t)$ where $\xi$ is arclength
and $t$ is time.
In the absence of friction the velocity $d{\bf s}/dt$
of the filament at the point ${\bf s}$ is given by\cite{Schwarz}

\begin{equation}
\frac{d{\bf s}}{dt}=
{{\Gamma}\over{4\pi}}
\int {{( {\bf r} - {\bf s} ) \times {\bf dr}}
\over{\vert  {\bf r} - {\bf s} \vert^3}},
\end{equation}

\noindent
where $\Gamma$ is the quantum of circulation and the Biot - Savart integral
extends over all vortex filaments present in the flow (we do not apply 
the local induction approximation).
By changing the discretization along the filaments we tested that the
results do not depend on the reconnection procedure (details of the
algorithm are in reference\cite{numerics}).
Since our model is incompressible, we have no 
transformation of kinetic energy (length of vortex line) into sound, 
an effect which can be studied using the Gross - Pitaevski 
model\cite{sound}. A small loss of vortex length occurs 
due to the numerical reconnection procedure\cite{Tsubota}
but does not affect our results.

Using the above model, we have performed numerical
experiments to determine
the spatial - temporal evolution of localized packets of vorticity.
Although we used $^4$He's parameter values, the results
can be reinterpreted for $^3$He by rescaling the units of time and length
according to the different values of $\Gamma$. The vortex filament
method, originally developed for $^4$He, is equally valid for
$^3$He-B despite the much larger vortex core size than $^4$He, 
because the length
scales of the calculation (eg distance between discretization
points on the same filament and distance between filaments) are
still much bigger than the vortex core.
Typically our initial condition consists of a given number $N_0$
of circular vortex rings, whose centres and orientations are
randomly generated, initially confined in a sphere of radius $S_0$.
Other kinds of initial conditions have been discussed in the
related literature\cite{init}, notably that of random vortex network,
but there is no reason to believe
that they apply to our case and we know too little of the details
of how a vibrating wire or grid generates quantized vorticity
to be more realistic. Fortunately it is known\cite{Schwarz}
that any simple configuration which is almost
isotropic quickly evolves in an turbulent tangle independent of the
initial state, and by numerically experimenting we found that our results
do not depend on how we start the calculation. For example we tried
replacing many small circular rings with few longer Fourier knots
(trefoil-like curves which wrap around themselves few
times before closing\cite{Kauffman}). This changed drastically the
initial topology, but the same results were found as for rings.
Unlike previous numerical simulations of superfluid turbulence (performed
either with periodic boundary conditions or in channels with rigid
walls), our calculations are carried out in an infinite volume.
At each time, quantities which are useful to describe the vortex packet are:
the total length $\Lambda$, the number of loops $N$,
the radius of the confining sphere $S$, the vortex line density 
$L=3 \Lambda/4 \pi S^3$, the average inter-vortex distance
$\delta=L^{-1/2}$, and the average vortex loop length
in the packet $D$. These quantities depend on $t$ and we
use the subscript zero to denote initial values.

The time evolution of the small vortex packet shown in Figure 1 is typical.
The initial vortex rings (here $N_0=20$) interact, become
distorted and reconnect.
The evolution of the vortex packet is determined by the balance between 
self-reconnections and reconnections between different loops\cite{Tsubota}.
During the initial coalescence phase the reconnections
between different loops dominate and the number of separate loops
decreases ($N(0.34)=12$ in Figure 1).
Following this comes an evaporation phase, when
self-reconnections dominate. In this phase 
occasionally a self-reconnection generates a small vortex
loop smaller than the average separation between the
vortices. Because of its small size, the loop moves relatively fast,
and, if created near the surface of the packet and oriented in the
correct direction, it escapes from the packet,
flying to infinity without further reconnections.
Since each loop which escapes decreases the total vortex
length in the packet, the average distance between
vortices increases, thus increasing the probability
that another loop escapes.
The evaporation and escape phases distinguish homogeneous from 
inhomogeneous turbulence.  In the former self-reconnections
and reconnections between different loops come to a balance, and a steady 
state tangle forms\cite{Schwarz}. In the latter, vortex
loops can quickly move out of the vortex packet and then never 
reconnect with other loops. A key ingredient of the effect
is therefore the counter-intuitive dispersion relation
of vortex loops: unlike particles, the less energetic (smaller)
they are, the faster they move.
At this point of the evolution the number of vortex loops become
constant (for example $N(2.2)=N(3.48)=37$ in Figure 1).
Self-reconnections come to dominate the isolated
vortex packet and the evaporation phase persists until the packet has
expanded away. The speed of an escaping loop can be
estimated from the well known formula for a classical
vortex rings of radius $R$

\begin{equation}
v_R=\frac{\Gamma}{4 \pi R} [\ln(8R/a)-1/2],
\end{equation}

\noindent
where $a$ is the vortex core radius.
In the actual experiments of course the volume is not infinite and
the motion of isolated loops which escape
from the packet terminates at the walls.

The interesting question is
what determines the characteristic timescale for the vortex packet
to evaporate. Since one important parameter of the problem is certainly
the quantum of circulation, $\Gamma$, to obtain a time scale
${\ell}^2/\Gamma$ we must identify the relevant length scale $\ell$.
There are only two length scales in the problem: the size of the
packet, $S$, and the average distance between the vortices, $\delta$,
which both change with time.  For the sake of simplicity, hereafter we refer
to the initial values $S_0$ and $\delta_0$.
The reason is that the definitions of $S$ and $\delta$
at later times are somewhat arbitrary, as they are
sensitive to the presence of small (fast) loops in a
particular numerical calculation. The use of the initial values simplifies
the analysis and lets us concentrate on the simple issue of whether
we can predict the evolution of the packet given initial
length $\Lambda_0$ and size $S_0$.

Figure 2 shows how a typical distribution of loop lengths
changes with time. At $t=0$ the distribution is on the $11^{th}$ bin
(all $N=30$ loops have the same length $d=0.067 \rm cm$ by
construction). As time proceeds the distribution
moves to the $3^{rd}$ bin (centred at $d\approx 0.013\rm cm$).
Note the direct cascade from the initial
peak at the right to the final peak at the left
without creation of intermediate length scales. The position
and height of the initial peak depends on the
initial configuration (if we have few longer Fourier knots at the
place of many small circular loops, the initial peak is smaller
and more to the right). What is universal is the
creation of the final left peak, which happens in all our simulations.

Now we analyze how the average loop length, $D=<~d~>$,
depends on $t$. For the sake of clarity, we normalize 
$D$ using the maximum value $D_{max}$ achieved in each
particular run. If we plot $D/D_{max}$ versus $t$ we note
an initial increase (coalescence) followed by a decrease
(evaporation) which eventually remains constant as separate loops 
fly away in all directions (escape).
The time scale for the packet to evaporate ranges in the interval
$0.05\rm sec < t < 2\rm sec$, depending on the particular run.
In Figure 3 we plot $D/D_{max}$
versus the scaled time $t/\tau$ where

\begin{equation}
\tau=\frac{\delta_0^2}{\Gamma}.
\end{equation}

\noindent
It is apparent that curves corresponding to the evolution of different
vortex packets now overlap, and evaporation takes place within the
shorter interval $1.5<t/\tau<2$. If we plotted the same graph by scaling
$t$ with $S_0$ rather than $\delta_0$ the curves would be very separate.
Figure 3 therefore suggests that the characteristic time scale of
evaporation is of the order of $\delta_0^2/\Gamma$.
Physically, $\tau$ represents the time scale of reconnections.
In fact, from the quantization of vorticity
$\oint {\bf v}_s \cdot {\bf dl}=\Gamma$, we estimate that
the typical speed inside the packet is of the order of
$v_s \approx \Gamma/\delta$ hence the typical reconnection time is
of the order of $\delta/v_s=\delta^2/\Gamma=\tau$. Note that,
since the distribution
of values of $\delta$ is large, some filaments reconnect earlier, which
is evident in Figure 1 at the beginning of the run.
The insert in Figure 3 shows the
evolution in space and time of different vortex packets.
Because of the above mentioned difficulty with the
definition of $S$, we use the more robust quantity $S'$, defined as the
radius of the sphere which contains {\it half} the total vortex length.
After the evaporation, the packet becomes a gas of loops which
fly to infinity, so we expect $S'\approx v t$ where $v$ is the speed
of the typical loop. From (2) we have, for $t >\tau$, 

\begin{equation}
\frac{S'}{\delta_0} \approx \Bigl(\frac{\cal L}{4 \pi}\Bigr)
\Bigl(\frac{t}{\tau}\Bigr),
\label{fit}
\end{equation}

\noindent
where $\cal L$ is a term with a weak logarithmic dependence on $\delta_0$.
The insert of Figure 3 confirms that $S'/\delta_0$ and $t/\tau$ are
proportional. The evolution of all
packets are similar and collapse onto the same curve, as shown by
the solid line which represents Eq.~(\ref{fit}).

We can now interpret the $^3$He-B turbulence experiments\cite{Fischer}.
In $^3$He we have $\Gamma=h/2m_3=6.6 \times 10^{-4}\rm cm^2/sec$
and $a\approx 10^{-6}\rm cm$. The tangle is created by
a vibrating NbTi filament bent into an approximately
semicircular shape with diameter $0.3\rm cm$, so we assume that the
initial vortex packet has dimension $S_0\approx 0.1\rm cm$. We note that
the time scale $S_0^2/\Gamma \approx 15\rm sec$ is far too large to have
relevance to what is observed. The number of vortices required to
produce the observed barrier to the quasi - particles is
estimated by the authors to correspond to a flow of order
$v_s \ge 0.1 \rm cm/sec$, hence, from $v_s=\Gamma/2\pi \delta_0$,
we estimate $\delta_0 \ge 10^{-3}\rm cm/sec$, and we conclude that
the vortex line density
must be of the order of $L_0 \le 10^6\rm cm^{-2}$. The characteristic
time scale for the vortex packet to evaporate in a gas of small rings
is therefore of the order of
$\tau=\delta_0^2/\Gamma \approx 1.5\times 10^{-3}\rm sec$. The small loops
fly away with speed $v_R \le 0.4\rm cm/sec$ estimated from equation (2)
since we know that $R \ge 10^{-3}\rm cm$. This result, that the velocity of
expansion of the quantized vorticity is of the order of $1\rm mm/sec$,
is consistent with the observation that the vortex tangle spreads over
the distance of $1\rm mm$ in the time of approximately $1\rm sec$.

In conclusion, we have shown that a packet of quantized vorticity,
initially localized in a small region of space, evaporates\cite{preprint} and
diffuses away as a gas of small vortex loops on the time scale
of order $\tau=1/L\Gamma$.
Application of
this scenario to the recent turbulent $^3$He-B experiment yields
order of magnitude agreement with the observed
spatio - temporal evolution.  This cascade to small loops is similar to
an idea
originally proposed by Feynman\cite{Feynman}. To pursue this study
in the context of $^4$He it would be interesting to use the 
Gross - Pitaevskii model
to determine whether small vortex loops radiate phonons.

Finally, our results should be of interest in fluid dynamics.
Firstly, we have found a peculiar form of
diffusion in what is actually an inviscid fluid. Secondly, we have
found a mechanism to transfer energy to small scales.
Thirdly, we have shown that, as far as helicity is concerned,
the superfluid represents a different, third benchmark to study, besides 
the traditional  Navier - Stokes and Euler fluids.

We thank S.N. Fisher and D.I. Bradley for discussions.

\noindent
Figure 1:

Evolution of a small vortex packet of quantized vorticity
(data corresponding to the crosses in Figure 3).
\medskip

\noindent
Figure 2:

Number $N$ of vortex loop of given length
at different times corresponding to the upward
triangles in Figure 3. Note the direct cascade from the right peak to the
left one (the small peak in the middle is due
to some larger loops left at the centre of the evaporated packet).
\medskip

\noindent
Figure 3:

Normalized average loop length $D/D_{max}$ vs $t/\tau$. 
The values of $N_0$, $S_0$ ($cm$), $L_0$ ($cm^{-2}$)
and $D_{max}$ ($cm$) for each run are:
Stars: $30$, $0.090$, $1252$ and $0.38$;
%
Crosses: $20$, $0.090$, $835$ and $0.22$;
%
White circles: $30$, $0.018$, $31296$ and $0.059$;
%
Upward triangles: $30$, $0.045$, $5007$ and $0.139$;
%
Black squares: $25$, $0.018$, $26080$ and $0.042$;
%
Downward triangles: $25$, $0.027$, $11590$ and $0.081$;
%
White squares: $60$, $0.090$, $2504$ and $=0.339$.
%
Black circles: $4$, $0.018$, $31460$ and $0.194$
(in this case the initial condition consists of few long Fourier
knots, so, unlike the other runs, reconnections immediately increase the
number of separate loops and $D$ is maximum at $t=0$).

The insert ($S'/\delta_0$ vs $t/\tau$) shows
how the evolution of different packets
scale together. The solid line shows Eq.~(\ref{fit}).
\bigskip

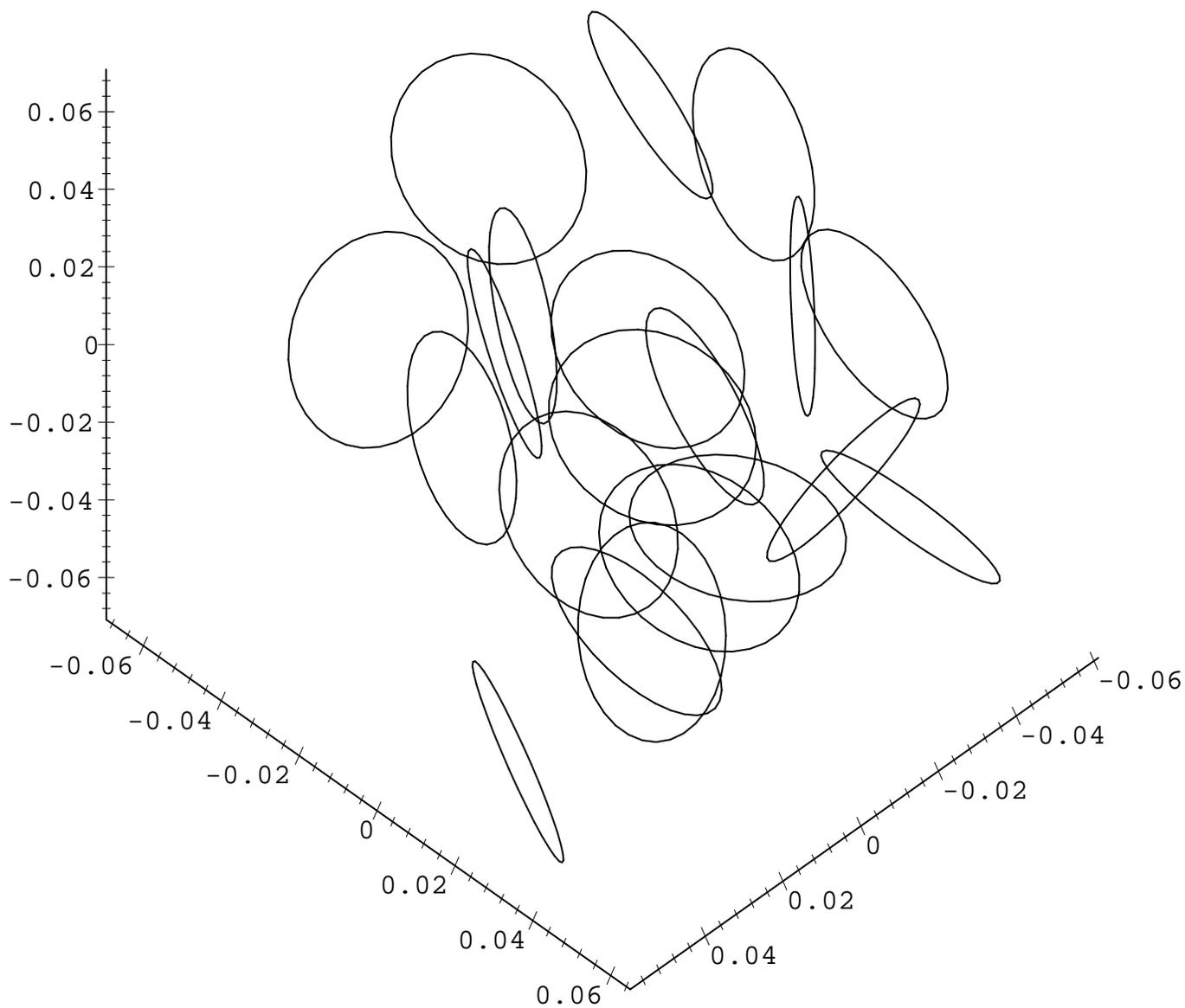
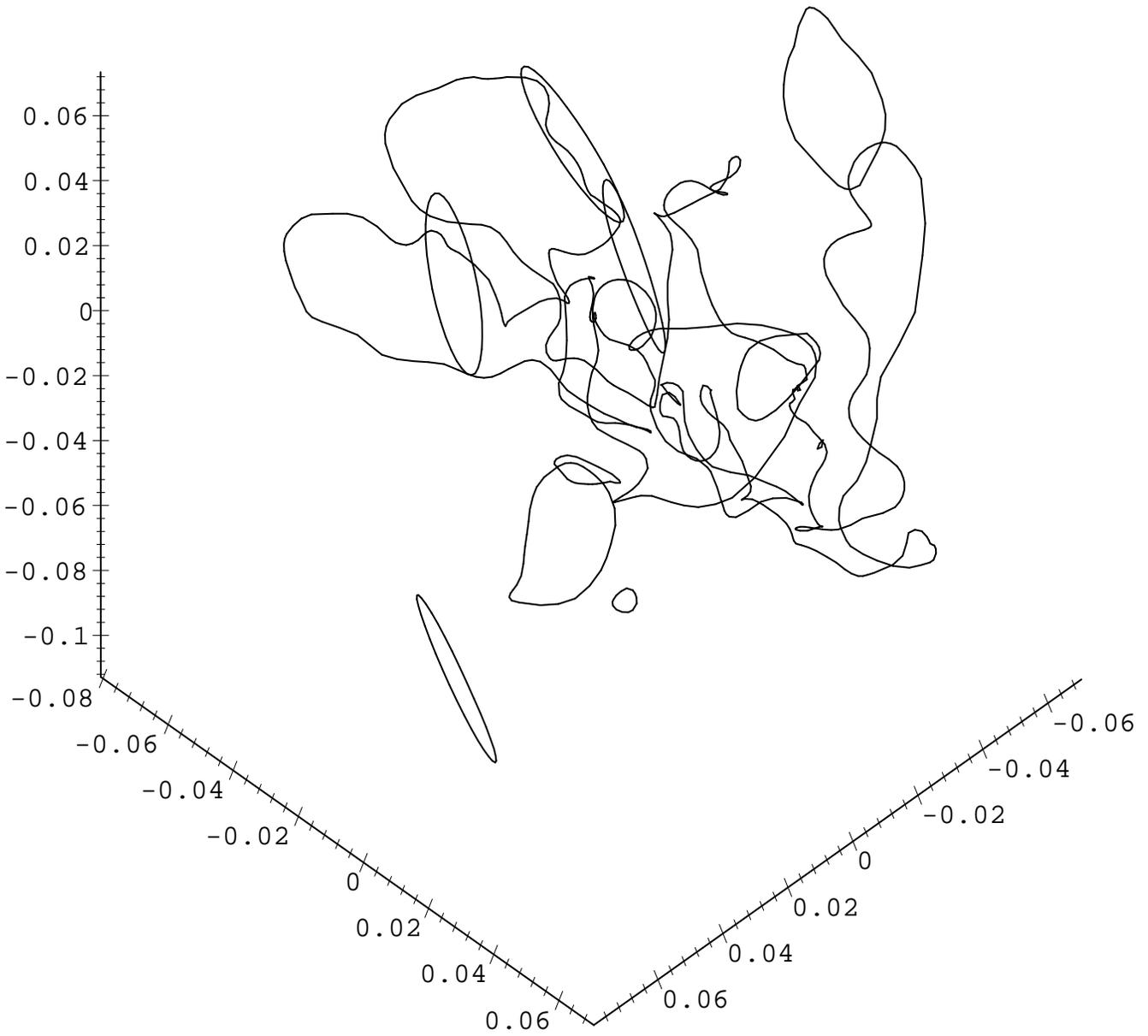
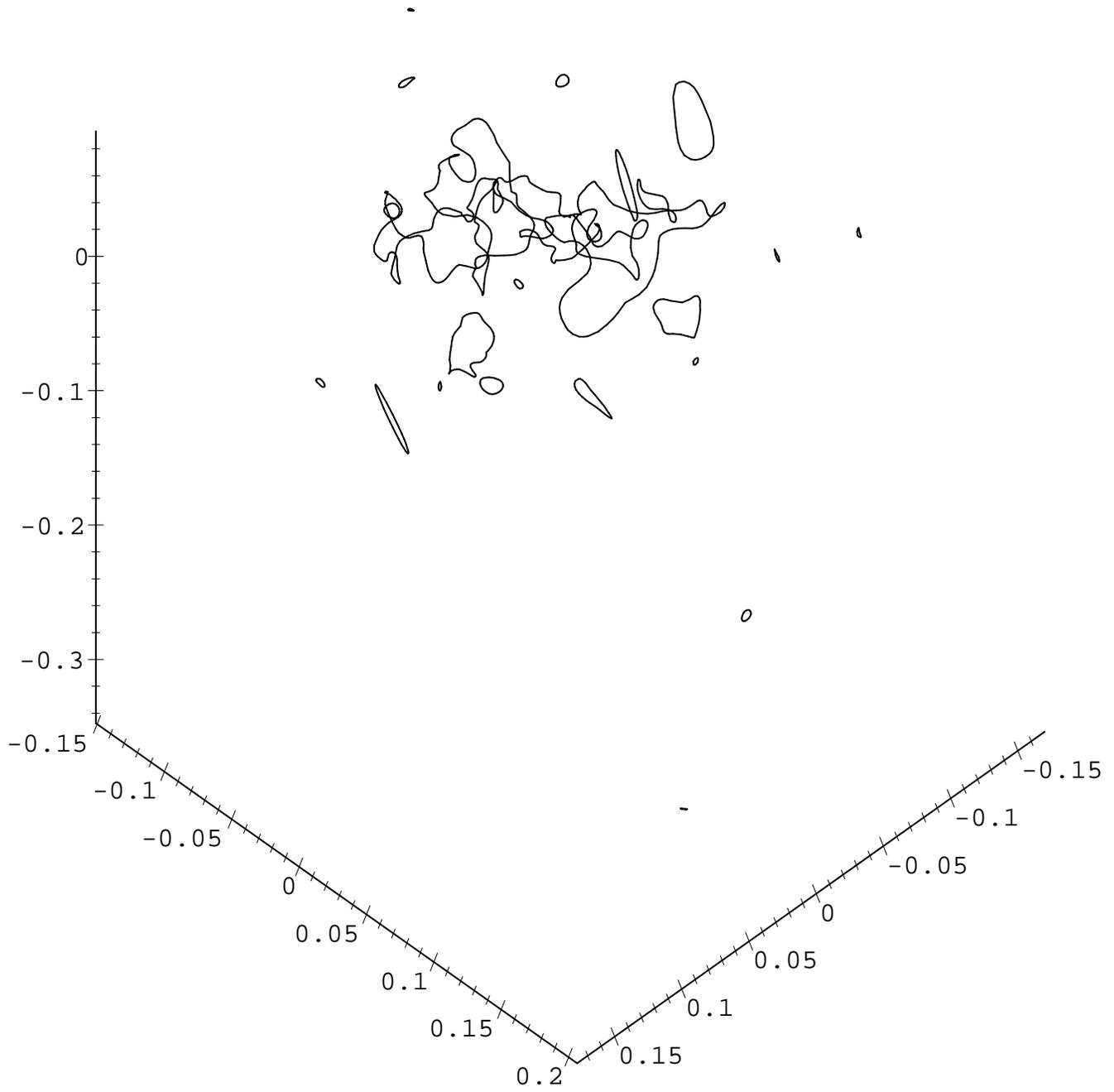
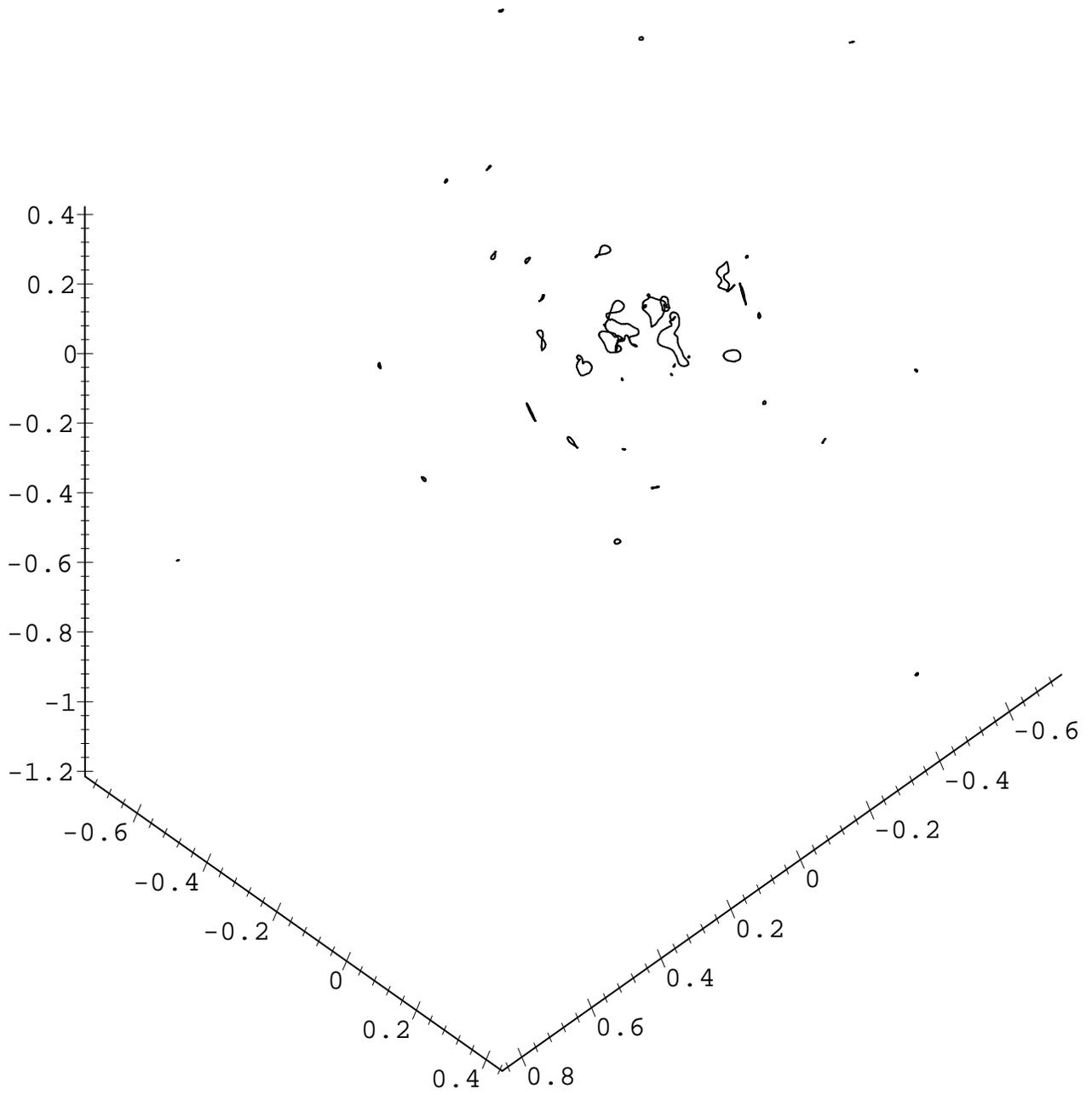
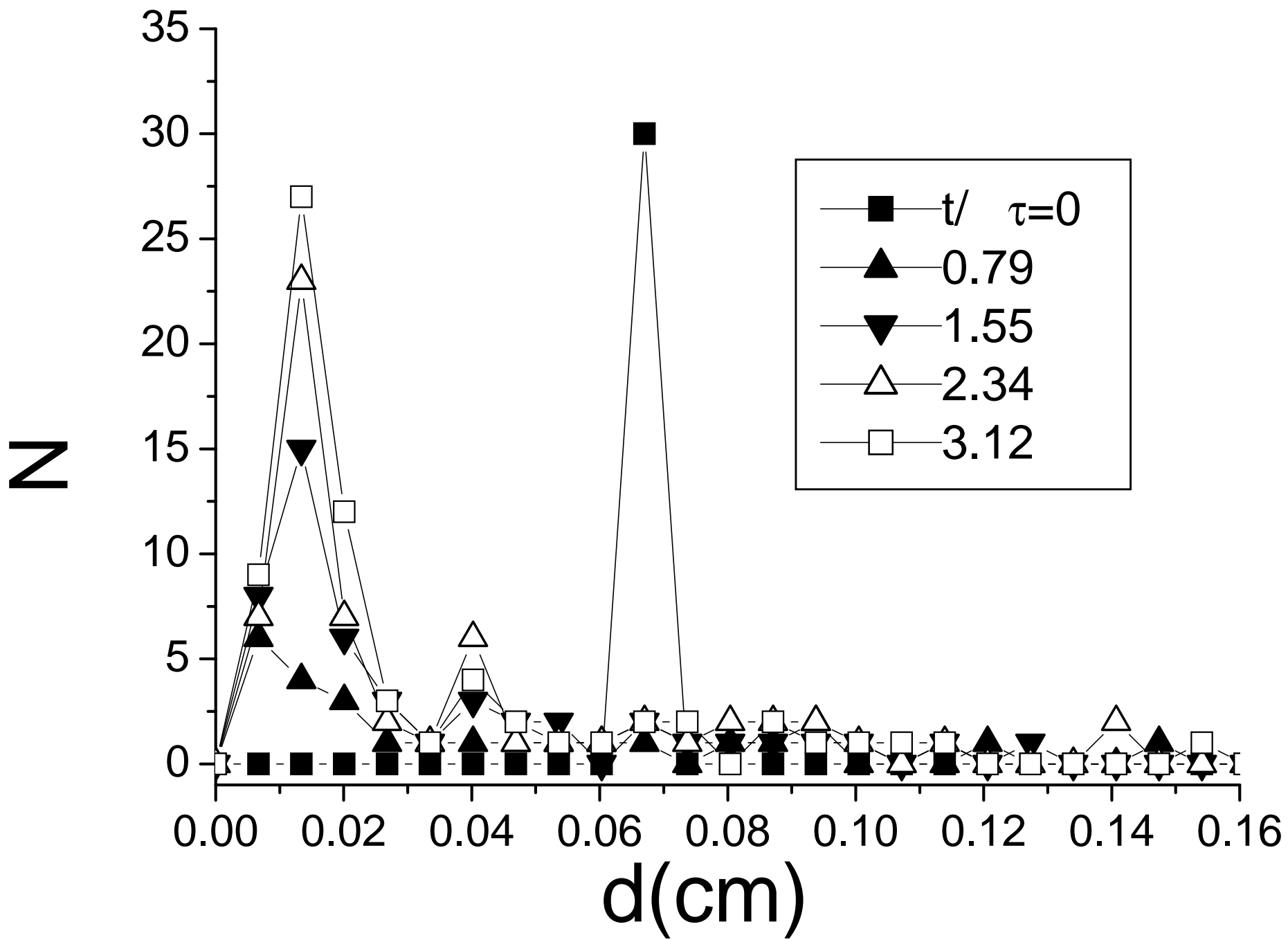
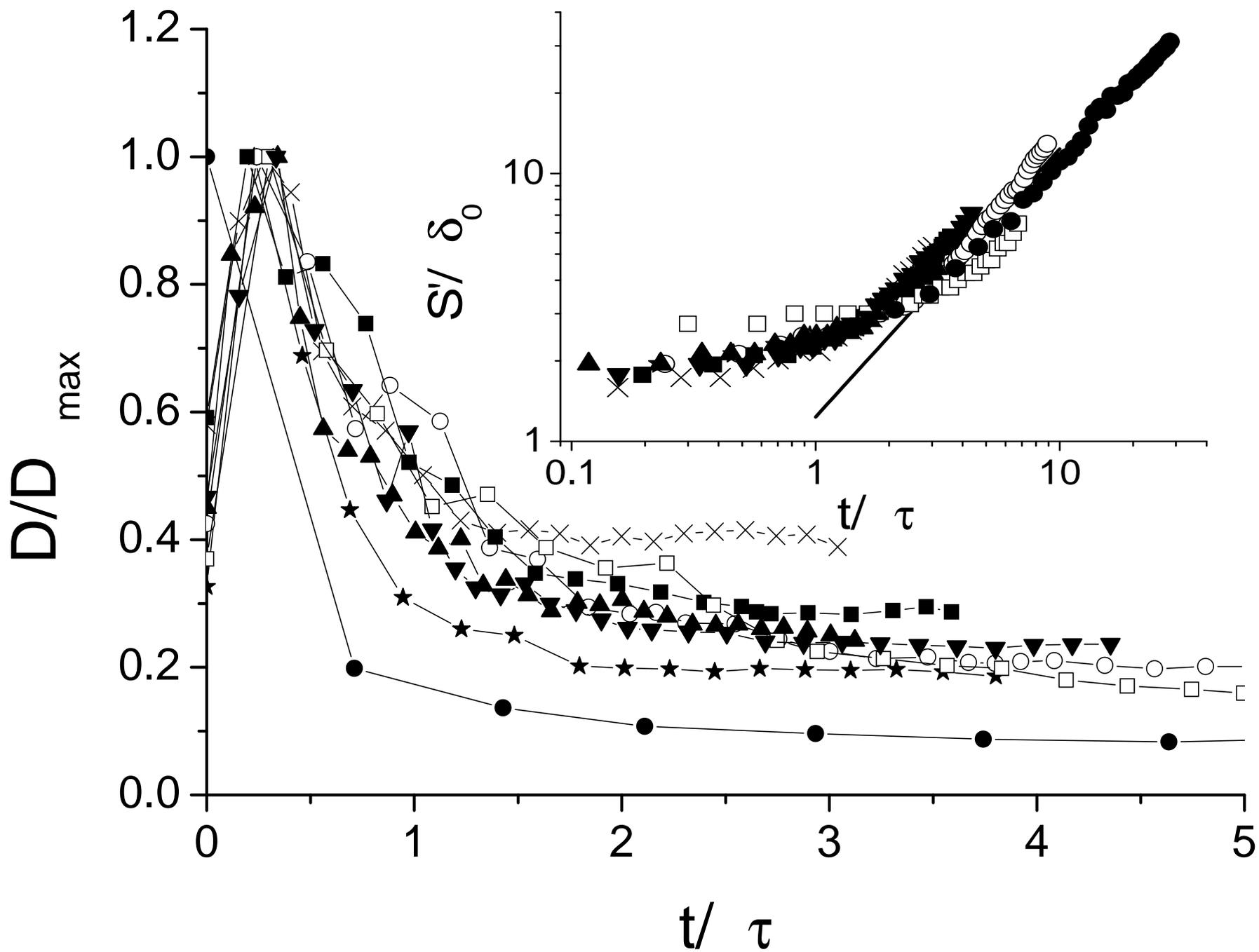
\end{document}